\documentclass{article}
\usepackage[utf8]{inputenc}
\usepackage{authblk}
\usepackage[dvips]{graphicx}
\graphicspath{{noiseimages/}}
\usepackage{xcolor}
\usepackage[T2A]{fontenc}
\usepackage{tikz}
\usepackage{pgfplots}
\usepackage{mathrsfs} 
\pgfplotsset{compat=1.7}
\usetikzlibrary{intersections, pgfplots.fillbetween}
\usetikzlibrary{decorations.pathmorphing}
\usetikzlibrary{arrows.meta}
\usepackage[mode=buildnew]{standalone}
\usepackage[toc,page]{appendix}
\usepackage{amsmath}
\usepackage{amssymb}
\usepackage{float}
\usepackage{comment}
\usepackage{a4wide}
\usepackage[english]{babel}
\usepackage{hyperref}
\definecolor{urlcolor}{HTML}{990000}
\definecolor{linkcolor}{HTML}{005F5F} 
\hypersetup{pdfstartview=FitH,  linkcolor=linkcolor,urlcolor=urlcolor, colorlinks=true,citecolor=blue}
\usepackage[page]{appendix}

\newcommand{\w}{\omega}

\author[1,2]{K.V.Bazarov \footnote{\tt bazarov.kv@phystech.edu}
}
\affil[1]{Moscow Institute of Physics and Technology, Institutskii per. 9, 141700, Dolgoprudny, Russia}
\affil[2]{ Institute for Theoretical and Experimental Physics, B. Cheremushkinskaya 25, 117218, Moscow, Russia}
\title{\textcolor{black}{Notes on peculiarities of quantum fields in space-times with horizons}}

\begin{document}

\numberwithin{equation}{section}

\maketitle

\begin{abstract}
We consider massive scalar field theory on static four-dimensional space-times with horizons. We  study the near horizon behaviour of the quantum expectation values of the stress--energy tensor operator for thermal state with generic temperatures. It turns out that the dependence of the expectation values on the temperature and tensor structure of the stress--energy tensor differs from the usual one in the Minkowski space-time. Moreover, for non--canonical temperatures these expectation values are divergent on the horizons. We also show that the Wightman functions have additional infrared peculiarities near the horizons.

\end{abstract}


\section{Introduction}

The main goal of this paper is to calculate the quantum expectation value of the stress--energy tensor (SET)\footnote{Here and after abbreviation "SET" stands for "quantum expectation value of the stress--energy tensor"} operator in several gravitational backgrounds with horizons. The reason why such calculations are interesting is that in the semiclassical approximation for gravitation, the backreaction problem is addressed via the solution to the Einstein equation of the following form \cite{birrell_davies_1982}:

\begin{align}
\label{ein}
    G_{\mu\nu}+\Lambda g_{\mu\nu}=8\pi G \langle :\hat{T}_{\mu\nu}:\rangle,
\end{align}
where $G_{\mu\nu}$ is the Einstein tensor and the expectation value in question appears on the right hand side (RHS).

Indeed, during the first step, pure gravity theory is usually considered, and only then one consider quantum fields, which live on the solution of the classical Einstein equations. Such calculation is based on the assumption, that the RHS of eq. \eqref{ein} may be neglected. On the one hand, this is a reasonable approximation, since the gravitational constant $G$ is a small parameter. On the other hand, we will see below, that discussing assumption breaks down in some cases. So direct calculation of the quantum expectation value of the stress--energy tensor makes it possible to estimate the value of the quantum fluctuation of the field to test the assumption under discussion.

The expectation value is taken over a state characterized by a density matrix $\rho$. In generic circumstances a quantum field may be in many different states, and there is no reason to restrict consideration to only one specific state \cite{Akhmedov:2021rhq}. However, in the Rindler space-time, one usually restricts consideration only to the so-called Minkowski vacuum or the Poincare invariant state (see, however, \cite{Akhmedov:2021agm}); in the de Sitter space-time — to isometry invariant states \cite{Bunch:1977sq}, \cite{Mottola:1984ar}, \cite{Allen:1985ux} (see, however, \cite{Akhmedov:2013vka} and \cite{Akhmedov:2019cfd}). Furthermore, in the Schwarzschild black hole background one considers only are the Boulware \cite{PhysRevD.11.1404}, Unruh \cite{PhysRevD.14.870} or Hartle-Hawking \cite{PhysRevD.21.2185,PhysRevD.13.2188} states (see, however, \cite{Akhmedov:2015xwa}). The Boulware state corresponds to the empty space for large radii $r$, while the Unruh state corresponds to the outgoing energy flux at the Hawking temperature. The Hartle-Hawking state corresponds to the surrounding a black hole quantum gas at the Hawking temperature, and SET of the latter state is regular. So, there are physical reasons to consider these three states. 
But, we find it appropriate to consider a wider range of states and calculate SET for these states in four-dimensional space-times with horizons. However, consideration of a generic reasonable state is a very hard problem. Hence, in this paper, we restrict our attention to the simple class of thermal states with generic temperatures. This class includes the isometry invariant states for the canonical values of temperature. But even though the states are thermal, we find a nontrivial result, similar to the one discovered in our previous papers \cite{Akhmedov:2021agm,Akhmedov:2020ryq,Diatlyk:2020nxa,Akhmedov:2020qxd, Anempodistov:2020oki}, which were devoted to the aspects of quantum fields physics in the background space-times with horizons. In those papers we mainly considered two-dimensional theories and studied the properties of the Wightman functions. In this paper we extend the consideration to four--dimensional theories and to the properties of the SET.

According to \cite{Kay:1988mu} KMS states with non-canonical temperature are singular in space-times with bifurcate horizons. However, that judgement is based on the properties of Wightman functions. But consideration of Wightman functions does not allow one to draw any conclusions about the physical consequences of such a singularity. In our paper we consider expectation value of stress-energy tensor and show that this singularity leads to strong backreaction of the quantum state on the background geometry. 

Roughly speaking, the physical question we would like to address in this paper is as follows: What is an appropriate way to consider a thermal gas of exact modes placed in a static space-time with a horizon? As such spaces, we consider three background metrics:

\begin{subequations}
\label{metrics}
\begin{align}
        ds^2= & e^{2\xi \alpha}\Big(d\eta^2-d\xi^2\Big)-d \vec{z}_{\perp}^2,     \label{metricRin} \\
        ds^2=f(r) \, dt^2  & -\frac{dr^2}{f(r)}-r^2\big(\sin^2 \theta d\varphi^2+\theta^2\big), \quad f(r)=1-H^2 \, r^2 \label{metricdS}\\
        ds^2=\Big(1-\frac{r_s}{r}\Big)dt^2  & -\frac{dr^2}{1-\frac{r_s}{r}}-r^2\big(\sin^2 \theta d\varphi^2+\theta^2\big), \quad r_s=2MG.\label{metricBH}
\end{align}
\end{subequations}
These three metrics describe the flat four-dimensional Rindler space, the static de Sitter space-time, and the Schwarzschild black hole, correspondingly.

Throughout the paper, we set the mass of the black hole to be $M=\frac{1}{4 G}$, the Hubble constant --- $H=1$, and the proper acceleration --- $a=1$. In such units, canonical temperatures (so called Hawking \cite{Hawking:1975vcx}, Unruh \cite{Unruh:1983ms}, and Gibbons-hawking \cite{Gibbons:1977mu} temperatures, correspondingly) are equal to $(2\pi)^{-1}$:
\begin{align}
\label{temps}
    \beta_H\equiv 8\pi G M=2\pi,\qquad \beta_U\equiv\frac{2\pi}{\alpha}=2\pi, \qquad \beta_{dS}\equiv\frac{2\pi}{H}=2\pi.
\end{align}
The setup of the problem is as follows: we consider the static background metric as the solution to equation \eqref{ein} for the corresponding value of the cosmological constant and without the SET on the RHS. Then on such a background, we quantize the massive scalar field with the following action:

\begin{align*}
    S = \frac{1}{2} \, \int d^4 x \, \sqrt{-g} \, \Big(\partial_\mu \varphi\partial_\mu \varphi-m^2 \varphi^2\Big).
\end{align*}
After that, we find the SET expectation value from the Wightman function as:
\begin{align}
\label{defT}
    T_{\mu\nu}(x)_\beta=\bigg(\frac{\partial}{\partial x_1^\mu}\frac{\partial}{\partial x_2^\nu}-\frac{1}{2}g_{\mu\nu}\Big[g^{\alpha\beta}\frac{\partial}{\partial x_1^\alpha}\frac{\partial}{\partial x_2^\beta}-m^2\Big]\bigg)W_\beta(x_1|x_2)\bigg|_{x_1=x_2=x},
\end{align}
which is defined as follows: 
\begin{align}
\label{defW}
    W_\beta(x_1|x_2)=\langle\hat{\varphi}(x_1)\hat{\varphi}(x_2)\rangle_\beta, \qquad \langle\hat{O}\rangle_\beta\equiv \frac{\text{Tr} e^{-\beta \hat{H}} \hat{O} }{\text{Tr} e^{-\beta \hat{H}}},
\end{align} 
for the temperature $T = 1/\beta$. The Hamiltonian operators have the standard form and are listed in the appendixes \eqref{hamrin}, \eqref{hamds}, \eqref{hamsch}. Evaluating the explicit expression of the SET for a generic state is a complex technical problem, as will be seen below. However, we can find its approximate value near the horizon for thermal state with generic temperature. That is precisely the part of space-time, which is of the main interest for us. 

Our work is based on an essential assumption that for the canonical temperature (\ref{temps}) the regularized SET near the horizon is zero up to terms, which lead to the cosmological constant renormalization. Let us clarify this point using an example of the Rindler and the de Sitter space-times. In these cases, the states with canonical temperature are Poincaré and de Sitter invariant, correspondingly (see, e.g., \cite{Akhmedov:2020qxd,Akhmedov:2020ryq}). Then the regularized SET should only renormalize the cosmological constant $\langle :\hat{T}_{\mu\nu}:\rangle \propto g_{\mu\nu}$ (These calculation was done using point-splitting methods see for details \cite{PhysRevD.13.2720,doi:10.1098/rspa.1977.0056,PhysRevD.18.4435,PhysRevD.17.946,PhysRevD.14.2490}). So we do not pay a lot of attention to it. Note, that in this paper we define the regularized expectation value of the stress-energy tensor as $\langle:\hat{T}_{\mu\nu}:\rangle$. This operation is not the same as an usual normal ordering.

We find it natural to use the following simplest scheme of regularization: we subtract the SET for the canonical temperature $\beta_C$ from the SET for a generic temperature $\beta$. In other words, we assume that the renormalized expectation value of the SET for a canonical temperature, $\langle:\hat{T}_{\mu\nu}:\rangle_{\beta_C}$, only renormalize cosmological constant and does not give any new physics:
\begin{align}
\label{regT}
   \langle:\hat{T}_{\mu\nu}:\rangle_\beta= \underbrace{\langle:\hat{T}_{\mu\nu}:\rangle_\beta-\langle:\hat{T}_{\mu\nu}:\rangle_{\beta_C}}_{\text{finite term(no regularization)}}+\langle:\hat{T}_{\mu\nu}:\rangle_{\beta_C}\approx\langle\hat{T}_{\mu\nu}\rangle_\beta-\langle\hat{T}_{\mu\nu}\rangle_{\beta_C},
\end{align}
Generally speaking, we are also interested in the tensor structure of the SET because, in cosmological models, one usually considers stress-energy tensors of the perfect fluid. It will be shown later that in our case, when the influence of the de Sitter space or the black hole is significant, the obtained SET is not described by an equation of the state of the perfect fluid. 

The paper is organized as follows. In Sec. \ref{Srind}, Sec. \ref{SdS}, Sec. \ref{SBH}, we calculate SET in the three different background space-times described above. Then in Sec. \ref{IRbeh}, we calculate the bahavior of the Wightman functions near the horizon. Some details of our numerical calculations are given in the appendixes.

\section{Rindler space-time \label{Srind}}

The four-dimensional Rindler space-time is described by the flat metric form eq. \eqref{metricRin}. This metric is obtained from the Minkowski one by the coordinates transformation. It is well-known, that such transformation has close connection to the quantization procedure \cite{Unruh:1983ms,PhysRevD.7.2850}. However, in this work we consider the metric \eqref{metricRin} as an independent object, meaning that if $\beta=2\pi$, then the Wightman function in the Rindler space-time coincides with the Wightman function of the Minkowski vacuum in the Minkowski space-time. In such a background,
the Klein-Gordon equation, and its solution can be written as:
\begin{align}
\label{eomRindler}
   \Big(-\partial_\xi^2+(m^2+k^2)e^{2\xi}\Big)f_{k,\w}(\xi)=\w^2f_{k,\w}(\xi), \quad \varphi(\eta,\xi,\vec{x})=e^{i\w t}e^{i\vec{k}\vec{x}}f_{k,\w}(\xi).
\end{align}
Then the field operator has the following mode expansion (see the details of quantization in \cite{birrell_davies_1982,Akhmedov:2020ryq}):
\begin{align}
\label{operatormassiveRindlerD}
    \hat{\varphi}(\eta,\xi,\vec{x}) =\int_{-\infty}^{+\infty} \frac{d^2k}{2\pi} \int_{0}^{+\infty} \frac{d \omega}{\pi}\sqrt{\sinh\pi \omega}\bigg[e^{-i\omega \eta+i\vec{k}\vec{z}}\hat{b}_{\omega,\vec{k}}^{}  +e^{i\omega \eta- i\vec{k}\vec{z}}\hat{b}_{\omega,\vec{k}}^\dagger\bigg]K_{i\omega}\big(\sqrt{m^2+k^2} e^{\xi}\big),
\end{align}
with the following commutation relations for the ladder operators:
\begin{align}
\label{commutaters}
    [\hat{b}_{\omega,\vec{k}}^{},\hat{b}_{\omega',\vec{k'}}^\dagger]=\delta(\w-\w')\delta^2(\vec{k}-\vec{k'}); \qquad [\hat{b}_{\omega,\vec{k}}^\dagger,\hat{b}_{\omega',\vec{k'}}^\dagger]=0; \qquad  [\hat{b}_{\omega,\vec{k}}^{},\hat{b}_{\omega',\vec{k'}}^{}]=0.
\end{align}
By definition \eqref{defW} and expression of the field operator \eqref{operatormassiveRindlerD}, the tree-level Wightman function is:
\begin{multline}
     W_\beta(\eta_1,\xi_1,\vec{z}_1|\eta_2,\xi_2,\vec{z}_2)=\langle\hat{\varphi}(\eta_1,\xi_1,\vec{z}_1)\hat{\varphi}(\eta_2,\xi_2,\vec{z}_2)\rangle_\beta=\\=\int_{-\infty}^{+\infty} \frac{d^2k d^2k'}{(2\pi)^2\pi^2} \int_0^{+\infty} d\omega d\omega'e^{i\vec{k} \vec{z}_1-i\vec{k}'\vec{z}_2} K_{i\omega}\big(\sqrt{m^2+k^2} e^{\xi_1}\big)K_{i\omega'}\big(\sqrt{m^2+k'^2} e^{\xi_2}\big) \times \\ \times   \sqrt{\sinh(\pi \omega)\sinh(\pi \omega')}\Big[e^{i \omega \eta_1-i \omega'\eta_2}\langle \hat{b}_{\omega,\vec{k}}^\dagger,\hat{b}_{\omega',\vec{k'}}^{} \rangle_\beta+e^{-i \omega \eta_1+i \omega'\eta_2}\langle \hat{b}_{\omega,\vec{k}}^{},\hat{b}_{\omega',\vec{k'}}^\dagger \rangle_\beta \Big].
\end{multline}
Quantum average of creation and annihilation operators is obtained from \eqref{defW}, where Hamiltonian has the form \eqref{hamrin}:
\begin{multline}
    \langle  \hat{b}_{\omega,\vec{k}}^\dagger,\hat{b}_{\omega',\vec{k'}}^{}  \rangle_\beta = \frac{\text{Tr} e^{-\beta \hat{H}}  \hat{b}_{\omega,\vec{k}}^\dagger,\hat{b}_{\omega',\vec{k'}}^{}  }{\text{Tr} e^{-\beta \hat{H}}}=\delta(\omega-\omega')\delta(\vec{k}-\vec{k}')\frac{\sum_{n=0}^{+\infty}e^{-\beta \omega n} n}{\sum_{n=0}^{+\infty}e^{-\beta \omega n} }=\\=\delta(\omega-\omega')\delta(\vec{k}-\vec{k}')\frac{1}{e^{\beta\omega}-1},
\end{multline}
and using \eqref{commutaters}:
\begin{align}
    \langle  \hat{b}_{\omega,\vec{k}}^{},\hat{b}_{\omega',\vec{k'}}^\dagger  \rangle_\beta=\delta(\omega-\omega')\delta(\vec{k}-\vec{k}')\Big(1+\frac{1}{e^{\beta\omega}-1}\Big)=-\delta(\omega-\omega')\delta(\vec{k}-\vec{k}')\frac{1}{e^{-\beta\omega}-1}.
\end{align}
Finally:
\begin{multline}
\label{Wbeta}
     W_\beta(\eta_1,\xi_1,\vec{z}_1|\eta_2,\xi_2,\vec{z}_2)=\\=\int_{-\infty}^{+\infty} \frac{d^2k d\omega}{(2\pi)^2\pi^2} \frac{\sinh(\pi \omega)}{e^{\beta \omega}-1}   e^{i \omega (\eta_1-\eta_2)}e^{i\vec{k} (\vec{z}_1-\vec{z}_2)} K_{i\omega}\big(\sqrt{m^2+k^2} e^{\xi_1}\big)K_{i\omega}\big(\sqrt{m^2+k^2} e^{\xi_2}\big).
\end{multline}
It may be worth stressing here again that we use the exact modes rather than plane waves. 

Now using this correlation function, we calculate the expectation value of the SET according to (\ref{defT}). The normal ordered energy density, defined as in \eqref{regT}, is as follows:
\begin{multline}
\label{modesT}
   \langle:\hat{T}_{00}(\xi):\rangle_{\beta}=\frac{1}{2}\int_{-\infty}^{+\infty}d\w\bigg[\frac{\sinh(\pi \omega)}{e^{\beta \omega}-1}-\frac{\sinh(\pi \omega)}{e^{2\pi \omega}-1} \bigg]  \times\\\times\int_0^\infty \frac{|k| d|k| }{(2\pi)\pi^2} \bigg[ \Big(\w^2+e^{2\xi}|k|^2\Big) \Big(K_{i\omega}\big(\sqrt{m^2+k^2} e^{\xi}\big)\Big)^2+\Big(\partial_\xi K_{i\omega}\big(\sqrt{m^2+k^2} e^{\xi}\big)\Big)^2\bigg].
\end{multline}
First, in the horizon limit $\xi\to-\infty$, for $|k| < \w e^{-\xi}\gg 1$, the Macdonald function can be replaced by a constant. On the other hand, in the opposite limit $|k|\gg\w e^{-\xi}$, the Macdonald function tends to zero as $K_{i\w}\Big(k e^{\xi}\Big)\sim\text{exp}(-ke^\xi)$, as $\xi\to\infty$. So, we see that there is an effective UV cutoff of the perpendicular momentum, i.e., $|k|<\w e^{-\xi}$. In other words, in the horizon limit, the leading contribution to the integral has the following form:
\begin{align}
\label{preint}
    \int_0^\infty d|k| |k|^n \Big(K_{i\omega}\big(\sqrt{m^2+k^2} e^{\xi}\big)\Big)^2 \propto \frac{1}{\w \sinh\pi \w} \int_0^{\w e^{-\xi}} d|k| |k|^n  \propto \frac{\w^{n+1} e^{-(n+1)\xi}}{\w \sinh\pi \w}.
\end{align}
Following this reasoning, we can approximate the integral over $|k|$ in \eqref{modesT} as $\frac{\w^4}{\sinh \pi \w}$. To the best of our knowledge, the explicit form of such integrals is not known. However, one can check numerically that the approximate values of the integrals in the limit $\xi\to-\infty$ are as follows:
\begin{align}
\label{I1}
    \int_0^\infty dk k\Big(K_{i\omega}\big(\sqrt{m^2+k^2} e^{\xi}\big)\Big)^2\approx\frac{\pi}{2}\w^2e^{-2\xi}\frac{1}{\w \sinh \pi \w},
\end{align}
\begin{align}
\label{I2}
    \int_0^\infty dk k^3 \Big(K_{i\omega}\big(\sqrt{m^2+k^2} e^{\xi}\big)\Big)^2\approx\frac{\pi}{3}\Big[\w^2+\w^4\Big]e^{-4\xi}\frac{1}{\w \sinh \pi \w},
\end{align}
\begin{align}
\label{I3}
    \int_0^\infty dk k \Big(\partial_\xi K_{i\omega}\big(\sqrt{m^2+k^2} e^{\xi}\big)\Big)^2\approx\frac{\pi}{6}\Big[4\w^2+\w^4\Big]e^{-2\xi}\frac{1}{\w \sinh \pi \w}.
\end{align}
The factor $\frac{1}{\w \sinh \pi \w}$ in \eqref{I1},\eqref{I2}, \eqref{I3} follows in the limit $\xi\to-\infty$ from the normalization of the Macdonald functions. Then, we consider the fixed $\w$ and calculate numerically the integrals over $k$ for some set of large negative values of $\xi$. This way, we find the dependence of \eqref{I1},\eqref{I2}, \eqref{I3} on $\xi$. We use the same idea to find the dependence of these integrals on $\w$. Then we consider the fixed large negative $\xi$, and calculate numerically the integrals over $k$ for some set of $\w$. One can do polynomial fit to find $\eqref{I1},\eqref{I2}, \eqref{I3}$. 

To make sure that our numerical methods work correctly, we estimate two of these integrals analytically in appendix \ref{appB}. That is done in the limit under discussion. We can do the analytical check in the Rindler space because we know the properties of the Macdonald functions in  \eqref{operatormassiveRindlerD} sufficiently well. However, in the cases of the black hole and the static de Sitter backgrounds, which are considered below, we have to deal only with the numerical estimates. Combining \eqref{I1}, \eqref{I2}, \eqref{I3}, and \eqref{modesT}, one obtains:
\begin{multline}
     \langle:\hat{T}_{00}(\xi):\rangle_{\beta}\approx \frac{e^{-2\xi}}{4\pi^2}\int_{-\infty}^{+\infty}d\w\bigg[\frac{1}{e^{\beta \omega}-1}-\frac{1}{e^{2\pi \omega}-1} \bigg] \Big(\w^3+\w\Big)=\\=\frac{e^{-2\xi}}{480\pi^2}\Big(\Big[\frac{2\pi}{\beta}\Big]^4-1\Big)+\frac{e^{-2\xi}}{48\pi^2}\Big(\Big[\frac{2\pi}{\beta}\Big]^2-1\Big), \quad \text{as} \quad \xi \to - \infty.
\end{multline}
The same way of reasoning can be applied for each other component of SET. Here we write down the results for all the components without derivation:
\begin{multline}
\label{genotv}
    \langle:\hat{T}^{\mu}_{\nu}:\rangle_{\beta=\frac{1}{T}}\approx\frac{1}{480\pi^2}e^{-4\xi}
\Big((2\pi T)^4-1\Big)\begin{pmatrix}
1 & 0 & 0 & 0\\
0 & -\frac13 & 0 & 0\\
0 & 0 & -\frac13 & 0\\
0 & 0 & 0 & -\frac13\\
\end{pmatrix}+\\+\frac{1}{48\pi^2}e^{-4\xi}
\Big((2\pi T)^2-1\Big)\begin{pmatrix}
1 & 0 & 0 & 0\\
0 & -\frac13 & 0 & 0\\
0 & 0 & \frac23 & 0\\
0 & 0 & 0 & \frac23\\
\end{pmatrix}, \quad \text{as} \quad \xi \to - \infty.
\end{multline}
This SET is not proportional to the metric tensor because the state that we consider here does violate the Poincare symmetry and describes matter on top of the isometry invariant state. Furthermore, one of its striking properties is that the SET infinitely grows near the horizon $\xi\to-\infty$. Meanwhile, the other side of the Einstein equations (\ref{ein}) is zero for the background under consideration. This means that the backreaction of the quantum fluctuation over the state under consideration on the background geometry cannot be neglected. Note that we consider here just the state with the Planckian (thermal) distribution for exact Ridler modes over the flat space-time.

Furthermore, the first term in \eqref{genotv} is similar to the standard SET of the perfect fluid. In the limit $T\to\infty$, we see the proper radiation-like behavior (up to the metric factor):
\begin{align}
  \lim_{T\to\infty}  \langle:\hat{T}^{\mu}_{\nu}:\rangle_{\beta=\frac{1}{T}}\approx \ \frac{T^4 \pi^2}{30}e^{-4\xi}
\begin{pmatrix}
1 & 0 & 0 & 0\\
0 & -\frac13 & 0 & 0\\
0 & 0 & -\frac13 & 0\\
0 & 0 & 0 & -\frac13\\
\end{pmatrix},
\end{align}
while in the opposite limit $T\to0$, we obtain:
\begin{align}
\label{Tto0}
  \lim_{T\to0}  \langle:\hat{T}^{\mu}_{\nu}:\rangle_{\beta=\frac{1}{T}}\approx \frac{1}{1440\pi^2}e^{-4\xi}
\begin{pmatrix}
-11 & 0 & 0 & 0\\
0 & 11 & 0 & 0\\
0 & 0 & -19 & 0\\
0 & 0 & 0 & -19\\
\end{pmatrix}.
\end{align}
At the same time, the canonical temperature, $T=\frac{1}{2\pi}$, corresponds to the Poincare invariant state, for which the properly normal ordered SET is $\langle:\hat{T}^{\mu}_{\nu}:\rangle_{2\pi}=0$, as it should be.

Let us stress that while the Poincare symmetry is broken, the general covariance remains intact:
\begin{align}
  D_\mu  \langle:\hat{T}^{\mu}_{\nu}:\rangle_{\beta}= \frac{\partial  \langle:\hat{T}^{\mu}_{\nu}:\rangle_{\beta}}{\partial x^{\mu}}+ \langle:\hat{T}^{\alpha}_{\nu}:\rangle_{\beta} \Gamma^\mu_{\alpha \mu}- \langle:\hat{T}^{\mu}_{\alpha}:\rangle_{\beta} \Gamma^\alpha_{\nu\mu}=0.
\end{align}
All in all, we encounter the situation as follows: while the background is flat Rindler space-time, and the state contains the planckian distribution for exact modes, the SET expectation value (\ref{genotv}) blows up near a horizon. This means that the back-reaction of the state under consideration on the geometry is strong and cannot be neglected. One has to solve Einstein equations with the expectation value of SET on the RHS. The situation is similar to the one encountered in \cite{Ho:2018jkm}.

\section{Four-dimensional static de Sitter space \label{SdS}}
\label{dssec}
In this section, we consider the curved de Sitter space-time in the static frame \cite{1917KNAB...19.1217D,desitter2,lem}. The metric of this space is \eqref{metricdS}. 
From now on, we set $H=1$.
The modes in such a background can be written in the following form:
\begin{align}
\label{fieldoperatordesitter}
    \varphi(t,r,\phi,\theta)=\frac{1}{\sqrt{4\pi \w}} \, \frac{e^{i\w t}}{r} \, Y_{l}^m(\phi,\theta)L(r^*), \quad \text{where} \quad   dr^*=\frac{dr}{f(r)}.
\end{align}
Then for the radial function, the Klein-Gordon equation acquires the form:
\begin{align}
\label{Heq}
    \Big[-\partial_{r^*}^2+V_l(r)\Big]L(r^*)=\w^2 L(r^*), \qquad V_l(r)=f(r)\Big(m^2+\frac{1}{r^2}l(l+1)+\frac{f'_r(r)}{r}\Big).
\end{align}
Thus, we obtain the Schrödinger-type equation. However, the potential $V_l(r)$ has peculiarity at the origin $r=0$. Namely, $V_l(r)\to\infty$ as $r\to 0\ (r^*\to-\infty)$. A general solution of the differential eq. \eqref{Heq} contains both growing and falling down solutions in the limit $r\to 0\ (r^*\to-\infty)$. However, on general physical grounds, we are interested only in the falling down (normalizable) modes as $r\to 0\ (r^*\to-\infty)$. We denote these modes as $L(r^*)$. At plus infinity $r^*\to+\infty$, they behave as linear combinations of two waves with the same amplitude propagating in opposite directions. As a result, the field operator has the following form:
\begin{align}
\label{fieldoperatords}
    \hat{\varphi}(t,r^*,\phi,\theta)=\sum_{l=0}^{\infty} \sum_{m=-l}^{l}\int_0^\infty d\w\frac{1}{r}\frac{1}{\sqrt{4\pi \w}}  \,\Big[ Y_{l}^m(\phi,\theta)e^{i\w t}L_{\w,l}(r^*)\hat{b}^\dagger_{\w,l}+h.c.\Big].
\end{align}
where $h.c.$ is hermitian conjugated term. By analogy of \eqref{Wbeta}, the two point Wightman function $W_\beta(x,x')$ is given by:
\begin{multline}
\label{Wds}
    W_\beta(x,x') = \Big\langle \hat{\varphi}(t,r,\theta,\phi)\varphi(t',r',\theta',\phi')\Big\rangle_\beta = \\ = \int_{-\infty}^{+\infty}d\w \frac{1}{e^{\beta \w}-1}\frac{e^{-i\w(t'-t)}}{rr'4\pi \w}\sum_{l=0}^\infty \sum_{m=-l}^l Y_l^m(\theta,\phi)\Big( Y_l^m(\theta',\phi')\Big)^*L_{\w,l}^*(r')L_{\w,l}(r)=\\=
    \int_{-\infty}^{+\infty}d\w \frac{1}{e^{\beta \w}-1}\frac{e^{-i\w(t'-t)}}{rr'4\pi \w}\sum_{l=0}^\infty \frac{2l+1}{4\pi}P_l\big(\vec{x}\cdot \vec{x}'\big)\Big[L_{\w,l}^*(r')L_{\w,l}(r)\Big],
\end{multline}
where $\vec{x}$ and $\vec{x'}$ lie on the unit sphere and $\vec{x}\cdot \vec{x}'=\cos\theta\cos\theta'\cos(\phi-\phi')+\sin\theta\sin\theta'$. To find the SET in the horizon limit, $r\to1$, we have to calculate the sum over $l$. Acting similarly to the previous section and using the numerical calculations (see appendix \ref{appA}) in the horizon limit, we obtain that:
\begin{subequations}
\begin{align}
\label{sumsds}
    \sum_{l=0}^{\infty}(2l+1) \big|L_{\w,l}(r)\big|^2\approx4\w^2 \Big(1-r^2\Big)^{-1},\\
    \sum_{l=0}^{\infty}(2l+1)l(l+1) \big|L_{\w,l}(r)\big|^2\approx \frac83 \Big(\w^2+\w^4\Big) \Big(1-r^2\Big)^{-2},\\
    \sum_{l=0}^{\infty}(2l+1) \big|\partial_{r^*}L_{\w,l}(r)\big|^2\approx \frac43 \Big(4\w^2+\w^4\Big) \Big(1-r^2\Big)^{-1}.
\end{align}
\end{subequations}
In the second sum, the factor $l(l+1)$ comes from the following expressions:
\begin{align}
   \partial_\theta\partial_{\theta'}P_l\big(\vec{x}\cdot \vec{x}'\big)\bigg|_{\vec{x}= \vec{x}'}=\frac{l(l+1)}{2}, \qquad \partial_\phi\partial_{\phi'} P_l\big(\vec{x}\cdot \vec{x}'\big)\bigg|_{\vec{x}= \vec{x}'}=\sin^2\theta \frac{l(l+1)}{2}.
\end{align}
Finally, the approximate form of the SET near the horizon is as follows:
\begin{multline}
\label{genotvds}
    \langle:\hat{T}^{\mu}_{\nu}:\rangle_{\beta=\frac{1}{T}}\approx\frac{1}{480\pi^2}(1-r^2)^{-2}
\Big((2\pi T)^4-1\Big)\begin{pmatrix}
1 & 0 & 0 & 0\\
0 & -\frac13 & 0 & 0\\
0 & 0 & -\frac13 & 0\\
0 & 0 & 0 & -\frac13\\
\end{pmatrix}+\\+\frac{1}{48\pi^2}(1-r^2)^{-2}
\Big((2\pi T)^2-1\Big)\begin{pmatrix}
1 & 0 & 0 & 0\\
0 & -\frac13 & 0 & 0\\
0 & 0 & \frac23 & 0\\
0 & 0 & 0 & \frac23\\
\end{pmatrix}.
\end{multline}
Again, we find that for generic values of the temprature $T$, the SET infinitely grows in the vicinity of the horizon. While for the canonical temperature, $T=1/2\pi$, it is vanishing. Thus, we encounter the same problem as in the Rindler space-time. Namely the backreaction of the thermal state with a generic temperature on the background geometry is strong.

\section{Schwarzschild black hole \label{SBH}}

Now we continue with the Schwarzschild black hole. The calculations in this section are similar to those in Sec. \ref{dssec}. As in the de Sitter space, we have to solve the Schrödinger equation. But, unlike the previous case, $V_l(r)$ is finite for the whole range of values of $r^*$. So, the full set of modes contains the out-going waves $R_{\w,l}(r)$ and the in-going waves $L_{\w,l}(r)$. The out-going ones behave as single waves at plus infinity $r^*\to\infty$, while the in-going ones behave as single waves at minus infinity $r^*\to-\infty$. So the field operator is given by:
\begin{align}
\label{fieldoperatorbh}
    \hat{\varphi}(t,r,\phi,\theta)=\sum_{l=0}^{\infty} \sum_{m=-l}^{l}\int_0^\infty d\w\frac{1}{r}\frac{1}{\sqrt{4\pi \w}}  \,\Big[ Y_{l}^m(\phi,\theta)e^{i\w t}\Big(R_{\w,l}\hat{a}^\dagger_{\w,l}+L_{\w,l}\hat{b}^\dagger_{\w,l}\Big)+h.c.\Big].
\end{align}
Note, that we have set of two creation and annihilation operators. The $a$ and $b$ operators denote the out-going and the in-going modes respectively. By analogy of \eqref{Wbeta} and \eqref{Wds}, the two point Wightman function $W_\beta(x,x')$ has the following form:
\begin{multline}
\label{Wbh}
    W_\beta(x,x')\equiv \langle \hat{\varphi}(t,r,\theta,\phi)\varphi(t',r',\theta',\phi')\rangle_\beta = \\ =\int_{-\infty}^{+\infty}d\w \frac{1}{e^{\beta \w}-1}\frac{e^{-i\w(t'-t)}}{rr'4\pi \w}\sum_{l=0}^\infty \sum_{m=-l}^l Y_l^m(\theta,\phi)\Big( Y_l^m(\theta',\phi')\Big)^*\times\\\times\Big[R_{\w,l}^*(r')R_{\w,l}(r)+L_{\w,l}^*(r')L_{\w,l}(r)\Big]=\\=
    \int_{-\infty}^{+\infty}d\w \frac{1}{e^{\beta \w}-1}\frac{e^{-i\w(t'-t)}}{rr'4\pi \w}\sum_{l=0}^\infty \frac{2l+1}{4\pi}P_l\big(\vec{x}\cdot \vec{x}'\big)\Big[R_{\w,l}^*(r')R_{\w,l}(r)+L_{\w,l}^*(r')L_{\w,l}(r)\Big].
\end{multline}
Our numerical calculations show that in the horizon limit, $r=\frac{1}{2}$, $(r^*\to-\infty)$, the main contribution comes from the out-going waves, while the in-going ones can be neglected. That is because the in-going waves have to tunnel through the potential $V_l(r)$ to reach the horizon. And this potential has a very high barrier for large values of $l$. 

Estimating the sums over $l$ in (\ref{Wbh}) gives (see appendix \ref{appA}):
\begin{subequations}
\label{sumsbh}
\begin{align}
    \sum_{l=0}^{\infty}(2l+1)\Big[ \big|L_{\w,l}(r)\big|^2+\big|R_{\w,l}(r)\big|^2\Big]\approx\sum_{l=0}^{\infty}(2l+1)\Big[ \big|R_{\w,l}(r)\big|^2\Big]&\approx\w^2 \Big(1-\frac{1}{2r}\Big)^{-1},\\
    \sum_{l=0}^{\infty}(2l+1)l(l+1)\Big[ \big|L_{\w,l}(r)\big|^2+\big|R_{\w,l}(r)\big|^2\Big]&\approx\frac{1}{6}\Big(\w^2+\w^4\Big) \Big(1-\frac{1}{2r}\Big)^{-2},\\
    \sum_{l=0}^{\infty}(2l+1)\Big[ \big|\partial_{r^*}L_{\w,l}(r)\big|^2+\big|\partial_{r^*}R_{\w,l}(r)\big|^2\Big]&\approx\frac{1}{3}\Big(4\w^2+\w^4\Big) \Big(1-\frac{1}{2r}\Big)^{-1}.
\end{align}
Then the approximate value of the SET near the horizon is as follows:
\begin{multline}
\label{genotvbh}
    \langle:\hat{T}^{\mu}_{\nu}:\rangle_{\beta=\frac{1}{T}}\approx\frac{1}{480\pi^2}\Big(1-\frac{1}{2r}\Big)^{-2}
\Big((2\pi T)^4-1\Big)\begin{pmatrix}
1 & 0 & 0 & 0\\
0 & -\frac13 & 0 & 0\\
0 & 0 & -\frac13 & 0\\
0 & 0 & 0 & -\frac13\\
\end{pmatrix}+\\+\frac{1}{48\pi^2}\Big(1-\frac{1}{2r}\Big)^{-2}
\Big((2\pi T)^2-1\Big)\begin{pmatrix}
1 & 0 & 0 & 0\\
0 & -\frac13 & 0 & 0\\
0 & 0 & \frac23 & 0\\
0 & 0 & 0 & \frac23\\
\end{pmatrix}.
\end{multline}
\end{subequations}
Similarly to the previous cases, for generic values of $T$, this SET infinitely grows near the horizon. We obtain results, which are close to \cite{doi:10.1080/00018738100101457}. But we consider an arbitrary temperature $\beta$. For the special case $\beta=\beta_H$, we obtain SET, which is regular at the horizon, as it was in \cite{doi:10.1080/00018738100101457}.

To conclude the last three sections, we see that for the three static frames with horizons under consideration, \eqref{metricRin}, \eqref{metricdS}, and \eqref{metricBH}, we obtain very similar results \eqref{genotv}, \eqref{genotvds}, and \eqref{genotvbh}, up to the constants $M,H$, and $\alpha$, which can be restored on dimensional grounds. Thus, one can see that if one considers a space-time with the metric of the form:
\begin{align}
\label{metricgen}
    ds^2=f(r)dt^2  & -\frac{dr^2}{f(r)}-r^2\big(\sin^2 \theta d\varphi^2+\theta^2\big),
\end{align}
where the function $f(r)$ has a first-order root at some point $f(r_0)=0$ with the canonical temperature $T_C$\footnote{Note that if $f(r)$ behaves near $r_0$ as: $f(r)\approx 2f_0 (r-r_0)$, then $T_C=\frac{f_0}{2\pi}$ is defined via the surface gravity.}. Then we expect that in the horizon limit ($r\to r_0$), the regularized SET has the following approximate form:
\begin{multline}
\label{genotvgen}
    \langle:\hat{T}^{\mu}_{\nu}:\rangle_{T}\underset{r\to r_0}{\approx}\frac{1}{480\pi^2}f(r)^{-2}
\Big((2\pi T)^4-(2\pi T_C)^4\Big)\begin{pmatrix}
1 & 0 & 0 & 0\\
0 & -\frac13 & 0 & 0\\
0 & 0 & -\frac13 & 0\\
0 & 0 & 0 & -\frac13\\
\end{pmatrix}+\\+\frac{1}{48\pi^2}f(r)^{-2}
(2\pi T_C)^2\Big((2\pi T)^2-(2\pi T_C)^2\Big)\begin{pmatrix}
1 & 0 & 0 & 0\\
0 & -\frac13 & 0 & 0\\
0 & 0 & \frac23 & 0\\
0 & 0 & 0 & \frac23\\
\end{pmatrix}.
\end{multline}
An important property of this expression is that it has two contributions. The first one depends on the temperature as $T^4$ and has the standard radiation type tensor form, while the second term is proportional to $T^2$ and is somewhat anomalous. 

It may be worth stressing here that we consider only the first-order roots in $f(r)$. Note that the second-order root scenario arises for an extremal black hole and, therefore, deserves special attention.

\section{IR behavior of the Wightman functions near the horizons}
\label{IRbeh}
This section will illustrate another interesting property of the scalar field theory in the three backgrounds mentioned above. The horizon is a light-like surface, so it is natural to expect that the Wightman function has the standard $UV$ divergence on the horizon. However, it was shown in \cite{Akhmedov:2020ryq,Akhmedov:2020qxd} that in the two-dimensional analogs of the spaces under consideration this divergence depends on the state of the theory over which the average in the correlation function is taken. Similar result in four dimensions was obtained in \cite{Haag:1984xa}. However, in \cite{Haag:1984xa} they have considered only the limit when two points points of the correlation function coincide. But in our paper we want to consider the behaviour of the Wightman function when the points are taken to the horizon in several different ways and do not necessary coincide. Also, according to the previous sections, our main interest are the expectation values of the stress-energy tensor rather than Wightman function.

Essentially, the observations of the previous paragraph mean that the $UV$ divergence depends on the $IR$ properties of the theory, which is rather odd. From the above discussion, we interpret this property as the strong backreaction of the quantum effects over the states under consideration on the background geometry. Still, in this section, we would like to separately examine the behavior of the Wightman functions near the horizons for the four-dimensional space-times under consideration. 

\subsection{Rindler space-time}

The limit in which two points of the Rindler space-time \eqref{metricRin} are simultaneously taken to the same part of the horizon is as follows:
\begin{align*}
    \xi_1=\xi_2\to -\infty,\quad \eta_2=\eta_1+\delta\eta,\quad \eta_{1,2}\to\infty,\quad 1\gg|\vec{z}_1-\vec{z}_2|\gg e^{\xi}.
\end{align*} 
In this limit, the Wightman function acquires the form:
\begin{multline}
\label{a21}
 W_{\beta}(\eta_1,\vec{z}_1,\xi|\eta_2,\vec{z}_2,\xi)=\int_{-\infty}^{+\infty} \frac{d^2k d\omega}{(2\pi)^2\pi^2} e^{i\vec{k}(\vec{z}_1-\vec{z}_2)} \frac{\sinh{\pi \w}}{e^{\beta \w}-1}e^{i \omega (\eta_1-\eta_2)} \bigg[K_{i\omega}\big(\sqrt{m^2+k^2} e^{\xi}\big)\bigg]^2=\\=
 \int_{-\infty}^{+\infty} \frac{ d\omega}{2\pi \pi^2}\frac{\sinh{\pi \w}}{e^{\beta \w}-1}e^{i \omega (\eta_1-\eta_2)} \int_0^\infty |k|d|k|J_0\big(|k||\vec{z}_1-\vec{z}_2|\big) \bigg[K_{i\omega}\big(\sqrt{m^2+|k|^2} e^{\xi}\big)\bigg]^2\approx\\\approx
 \frac{1}{|\vec{z}_1-\vec{z}_2|^2} \int_{-\infty}^{+\infty}\frac{ d\omega}{2\pi \pi^2}\frac{\sinh{\pi \w}}{e^{\beta \w}-1}e^{i \omega (\eta_1-\eta_2)} \int_0^\infty |k|d|k|J_0\big(|k|\big) \bigg[K_{i\omega}\Big(|k| \frac{e^{\xi}}{|\vec{z}_1-\vec{z}_2|}\Big)\bigg]^2.
\end{multline}
Using \cite{Gradshteyn:1702455}, one can calculate the following integral:
\begin{align*}
    \int_0^\infty x \Big[K_\mu(a x)\Big]^2 J_\nu(b x)dx=\frac{2 e^{2\mu\pi i}\Gamma(1+\frac\nu2+\mu)}{b(4a^2+b^2)^\frac12\Gamma(\frac\nu2-\mu)}Q_{\frac\nu2}^{-\mu}\Big(\sqrt{1+4a^2b^{-2}}\Big)Q_{\frac\nu2-1}^{-\mu}\Big(\sqrt{1+4a^2b^{-2}}\Big).
\end{align*}
Then in the limit under consideration $(1 \gg |\vec{z}_1-\vec{z}_2| \gg e^{\xi})$:
\begin{align*}
    \int_0^\infty |k|d|k|J_0\big(|k|\big) \bigg[K_{i\omega}\Big(|k| \frac{e^{\xi}}{|\vec{z}_1-\vec{z}_2|}\Big)\bigg]^2\approx-\frac{\pi}{\sinh\pi\w}\sin\bigg[2\w\log\bigg(\frac{e^{\xi}}{|\vec{z}_1-\vec{z}_2|}\bigg)\bigg].
\end{align*}
And finally:
\begin{multline}
\label{wbhorrin}
 W_{\beta}(\eta_1,\vec{z}_1,\xi|\eta_2,\vec{z}_2,\xi)\approx
 -\frac{1}{|\vec{z}_1-\vec{z}_2|^2} \int_{-\infty}^{+\infty}\frac{ d\omega}{2 \pi^2}\frac{1}{e^{\beta \w}-1}e^{i \omega (\eta_1-\eta_2)}\sin\bigg[2\w\log\bigg(\frac{e^{\xi}}{|\vec{z}_1-\vec{z}_2|}\bigg)\bigg] \approx \\ \approx \frac{1}{4\pi^2}\frac{1}{|\vec{z}_1-\vec{z}_2|^2}\frac{2\pi}{\beta},
\end{multline}
where in the last line, we take into account the contribution of only one pole under the integral, which is the closest to the real axis of the complex $\omega$-plane. Please note that we have shifted by $-i0$ the pole of the integrand in (\ref{wbhorrin}) in the complex $\w$-plane. The pole is due to the thermal distribution standing under the integral. This shift is done to obtain the proper zero temperature limit:
\begin{align*}
    \langle \hat{b}^\dagger_k \hat{b}^{}_{k'}\rangle=0, \qquad \langle \hat{b}^{}_k \hat{b}^\dagger_{k'}\rangle=\delta(k-k').
\end{align*}
To clarify this moment, let us consider a zero-temperature limit, i.e. $\beta\w\gg1$. If the pole at $\w=0$ in the thermal distribution is not shifted the contour of integration passes through it. But near the point $\w=0$ we have that $\beta\w\ll1$ for any $\beta$. Thus, to obtain the proper zero temperature behavior of the Wightman function, we have to shift the pole by $-i0$.

Above we have considered the following limit: $1\gg|\vec{z}_1-\vec{z}_2|\gg e^{\xi}$. Now we consider another limit $1 \gg e^{\xi}\gg|\vec{z}_1-\vec{z}_2|$. We set $\vec{z}_1=\vec{z}_2$ and $\xi_1=\xi_2\to-\infty$, then:
\begin{align}
 W_{\beta}(\eta_1,\xi|\eta_2,\xi)=\int_{-\infty}^{+\infty} \frac{d^2k d\omega}{(2\pi)^2\pi^2} \frac{\sinh{\pi \w}}{e^{\beta \w}-1}e^{i \omega (\eta_1-\eta_2)} \bigg(K_{i\omega}\big(\sqrt{m^2+k^2} e^{\xi}\big)\bigg)^2.
\end{align}
In the horizon limit $(\xi\to-\infty)$ using \eqref{I1}, one can find that:
\begin{align}
\label{IRnotHRind}
    W_{\beta}(\eta_1,\xi|\eta_2,\xi)\approx \frac{e^{-2\xi}}{4\pi^2}\int_{-\infty}^{+\infty}d\w \frac{\w}{e^{\beta \w}-1}e^{i \omega (\eta_1-\eta_2)}=\frac{e^{-2\xi}}{16\pi^2}\bigg(\frac{2\pi}{\beta}\bigg)^2\frac{1}{\sinh \Big(\frac{\pi(\eta_1-\eta_2)}{\beta}\Big)^2}
\end{align}
The key point of eq. \eqref{IRnotHRind} is as follows: in the limit under consideration the Wightman function is singular because of the factor $e^{-2\xi}$, which is divergent on the horizon. But in such a limit the functional dependence of the Wightman function on $\eta_{12}$ and $\beta$ is more complicated than it was above. Note that such a phenomenon does not occur in the two dimensional case.

Finally, the semi-UV behaviour in (\ref{wbhorrin}) has a wrong coefficient for non-canonical value of the temperature, $\beta$. Which means that the state under consideration is not of the Hadamard type. Yet this seems to be just a simple thermal state --- plankian distribution for the exact modes in the background under consideration. And let us stress that the leading contribution to (\ref{wbhorrin}) comes from the pole $\w=0$, which shows that the effect under consideration is the $IR$ one.
\subsection{Static de Sitter space}
In the Rindler space-time we have used properties of the Macdonald functions to calculate the behavior of the Wightman functions near the horizon. In the static de Sitter space-time the explicit form of the modes \cite{Polarski:1990ux} has a more complex structure. Because of that, instead of the mode expansion we use a trick as follows. If we know the Wightman function for the canonical temperature $\beta=2\pi$, then we also can find its form for the set of inverse temperatures of the form $\beta=2\pi/N$ with integer $N$ (for details see \cite{Akhmedov:2020ryq,Akhmedov:2020qxd,Akhmedov:2019esv}). 

In fact, the Wightman function with the canonical temperature $\beta=2\pi$ respects the de Sitter space isometry and has the following form \cite{Bunch,tagirov} (see also \cite{PhysRevLett.73.1746,Akhmedov:2019esv} for the recent discussion):
\begin{align}
\label{w2pids}
    W_{2\pi}(t_2-t_1,r_1,r_2,\vec{x}_1\cdot \vec{x}_2)=\frac{\Gamma\big(\frac{3}{2}+i\mu\big)\Gamma\big(\frac{3}{2}-i\mu\big)}{8\pi^2}\Big(Z^2-1\Big)^{-\frac12}P^{-1}_{-\frac12+i\mu}\Big(Z\Big), \qquad \mu\equiv\sqrt{m^2-\frac94},
\end{align}
where $Z$ is the hyperbolic distance between the two arguments $(t_1,r_1,\vec{x}_1)$ and $(t_2,r_2,\vec{x}_2)$ of the function:
\begin{align}
\label{hypd}
    Z=-\cosh(L)=-\sqrt{1-r_1^2}\sqrt{1-r_2^2}\cosh(t_2-t_1)-r_1r_2(\vec{x}_1\cdot \vec{x}_2),\qquad \vec{x}_{1,2}\in S_2.
\end{align}
This is just a hyperbolic function of the geodesic distance.
Near the horizon $(r_1=r_2=r\to1,(\vec{x}_1\cdot \vec{x}_2) \approx 1)$, \eqref{hypd} the hyperbolic distance behaves as:
\begin{align}
    Z\approx-(\vec{x}_1\cdot \vec{x}_2) \to -1.
\end{align}
Then, using the limiting form of the Legendre function \cite{bateman1953higher}, one can find the behavior of the Wightman function \eqref{w2pids} with the canonical temperature near the horizon:
\begin{align}
\label{WightasN}
    W_{2\pi}(t_2-t_1,r_1\approx1,r_2\approx1,\vec{x}_1\cdot \vec{x}_2)\approx\frac{1}{8\pi^2}\frac{1}{1-(\vec{x}_1\cdot \vec{x}_2)}.
\end{align}
Note that the obtained limiting expression does not depend on the time variables if $t_2-t_1$ remains finite in the horizon limit. 

Similarly, using the trick of \cite{Akhmedov:2020ryq,Akhmedov:2020qxd,Akhmedov:2019esv}, one can find that the Wightman function for the inverse temperature of the form $\beta=2\pi/N$ behaves near horizon as:
\begin{align*}
    W_{\frac{2\pi}{N}}\Big(t_2-t_1,...\Big)=\sum_{k=1}^N  W_{2\pi}\Big(t_2-t_1+2\pi i\frac{k-1}{N},...\Big)\approx N\times\frac{1}{8\pi^2}\frac{1}{1-(\vec{x}_1\cdot \vec{x}_2)}.
\end{align*}
Analytically continuing this answer to the non-integer $N$ of the form $N=\frac{2\pi}{\beta}$, we obtain the behavior of the Wightman function near the horizon for the general values of the temperature:
\begin{align}
\label{wbhords}
    W_\beta\approx \frac{2\pi}{\beta}\times\frac{1}{8\pi^2}\frac{1}{1-(\vec{x}_1\cdot \vec{x}_2)}.
\end{align}
This expression has a similar form to \eqref{wbhorrin} and again depends on the $IR$ properties of the theory.

As in the previous section, let us consider another limit in which first $\vec{x}_1=\vec{x}_2$ and then $r_1=r_2\equiv r\to 1$. In such a case:
\begin{align*}
    Z\approx-(1-r^2)\cosh(t_2-t_1)-r^2.
\end{align*}
Then, using \eqref{WightasN}, one can find that:
\begin{align}
    W_{\frac{2\pi}{N}}\Big(t_2-t_1,r,\vec{x}_1=\vec{x}_2\Big)\approx\frac{1}{1-r^2}\frac{1}{16\pi^2}N^2\frac{1}{\sinh \Big(\frac{N(t_1-t_2)}{2}\Big)^2}.
\end{align}
This equation has the same structure as \eqref{IRnotHRind}. Thus, the thermal state under consideration is not of the Hadamard type for non-canonical temperatures. 

\subsection{Black hole case}

Let us continue with the case of the black hole. The horizon limit should be taken in two steps. First, the radial coordinates of both points of the correlation function should strive to the horizon, i.e., $r_1^*=r_2^*=r^*\to-\infty$. Second, the angular coordinates of the two points should be set such that the two points would lie close to each other on the unit sphere, or $\theta\to0$. Moreover, we have to take $e^{r^*}\ll\theta$. This is just a limit similar to those we have considered above in the Rindler and static de Sitter space-times. In this limit, the following relation can be established numerically (see appendix \ref{appA}):
\begin{align}
\label{sumsbh2}
    \sum_{l=0}^{+\infty}\big(2l+1\big)P_l\big(\cos\theta\big)\Big[|R_{\w,l}(r^*)|^2+|L_{\w,l}(r^*)|^2\Big]\approx  \frac{8\w}{\theta^2}\sin\Big(2\w\log\frac{e^{r^*}}{\theta}\Big), \ \ \text{if} \ \ e^{r^*}\ll\theta\ll1.
\end{align}
Then the approximate form of the Wightman function \eqref{Wbh} is as follows:
\begin{align*}
    W_{\beta}\approx
    \int_{-\infty}^{+\infty}d\w \frac{1}{e^{\beta \w}-1}\frac{e^{-i\w(t'-t)}}{4\pi^2 \w}\frac{8\w}{\theta^2}\sin\Big(2\w\log\frac{e^{r^*}}{\theta}\Big)\approx\frac{2}{\pi}\frac{1}{\beta}\frac{1}{\theta^2}.
\end{align*}
Note that in the limit in question, the geodesic distance can be expressed as 
\begin{align*}
    \sqrt{L}\approx r\sqrt{2(1-\cos\theta)}\approx\frac{\theta}{2},
\end{align*}
because in the units that we use, the horizon corresponds to $r=\frac{1}{2}$. Finally, for the Wightman function, we obtain:
\begin{align}
     W_{\beta}\approx\frac{2\pi}{\beta}\times\frac{1}{4\pi^2}\frac{1}{L}.
\end{align}
In the limit $r_1=r_2\equiv r \to\frac12$ and $e^{r^*}\ll\theta$, using \eqref{sumsbh}, one obtains that:
\begin{align}
W_{\beta}(t_1,r|t_2,r) \approx \frac{\Big(1-\frac{1}{2r}\Big)^{-1}}{4\pi^2}\int_{-\infty}^{+\infty}d\w \frac{\w}{e^{\beta \w}-1}e^{i \omega (t_1-t_2)}=\frac{\Big(1-\frac{1}{2r}\Big)^{-1}}{16\pi^2}\bigg(\frac{2\pi}{\beta}\bigg)^2\frac{1}{\sinh \Big(\frac{\pi(t_1-t_2)}{\beta}\Big)^2}  . 
\end{align}
And again we find the same structure as in eq. \eqref{IRnotHRind}. Also we see that the behavior of the correlation function in the horizon limit depends on the IR properties of the theory. And, as in the previous cases, only the canonical Hawking temperature $\beta=2\pi$ gives the correct coefficient in front of the singularity at $L\to0$.

To conclude, let us point out that, interestingly enough, in all three cases considered in this section the singular near horizon contribution to the Wightman function has nothing to do with the singular behaviour of the SET encountered in the previous sections. This contribution just cancels out from the derivative part, $\partial \varphi \, \partial \varphi$, of the expression of the SET in the horizon limit. While the part of SET coming from the massive term, $m^2 \varphi^2$, is as usual suppressed in the horizon limit. 

\section{Conclusions and Acknowledgment}

We considered the near horizon behavior of the renormalized expectation value of the SET for the four-dimensional massive scalar field in the three static space-times with horizons: Rindler, de Sitter, and Schwarzschild. We also considered the behaviour of the Wightman function when its both points are light--like separated on the horizon. All these considerations are done in four dimensions, which extends the discussion of properties of the Wightman functions, which is done in earlier works in two dimensions. 

Here we consider a class of states described by the planckian density matrix for the exact modes. We refer to these states as thermal, although they do not possess all the necessary properties of the thermal state in the Minkowski space-time (for plane waves). That is one of the essential observations made in the present paper.

We made the following observations for all three space--times:
\begin{itemize}
	\item The SET expectation value infinitely grows near the horizon. It means that the back-reaction of a quantum fluctuation in \eqref{ein} is not negligible for the discussing set of states, and the background geometry is strongly affected.
	\item The expectation value is position dependent and has an unusual dependence on the temperature, which, in turn, signals an unusual equation of state.
	\item The dependence on the temperature and the tensor structure of the SET expectation value near horizon is the same for all three space-times. 
	\item We find that the behavior of the Wightman functions depends on the $IR$ properties of the theory when its both points are light--like separated on the horizons. This fact signals that these states are not thermal in the sense that we intuitively understand thermal states in the Minkowski space-time.
\end{itemize}
The key conclusion is that in curved backgrounds (and curved coordinates in flat space-time) for the states with non-canonical temperature (for thermal distribution for exact modes), it is necessary to consider the backreaction. This is quite unusual as compared to the thermal states for plane waves in Minkowski space-time.

I would like to acknowledge valuable discussions with O.Diatlyk, P.A.Anempodistov and D.V. Diakonov. I would like to thank E.T.Akhmedov for sharing his ideas and correcting the text. Also, I would like to thank E.M.Bazanova for proofreading the text of the paper. And I'm grateful to the referees for their constructive input. This work was supported by the grant from the Foundation for the Advancement of Theoretical Physics and Mathematics ``BASIS'' and by the Russian Ministry of education and science.
\begin{appendices} 
\numberwithin{equation}{section}
\setcounter{equation}{0}
\renewcommand\theequation{A.\arabic{equation}}

\section{Calculation of the Hamiltonians \label{appС}}

The Hamiltonian operator is defined as integration of the energy density of the space volume:
\begin{align}
\label{genham}
    \hat{H}=\int \sqrt{g}dx \hat{T}_0^0.
\end{align}
In the Rindler space-time, the calculation can be done straightforward. Using equation of motion \eqref{eomRindler}, one can rewrite massive term and spatial derivatives in \eqref{genham} in the following way:
\begin{align}
    \hat{H}=\int \sqrt{g}dx \hat{T}_0^0=\frac12 \int d\xi d^2\vec{z}\Big[\dot{\hat{\varphi}}(\eta,\xi,\vec{x})\dot{\hat{\varphi}}(\eta,\xi,\vec{x})-\hat{\varphi}(\eta,\xi,\vec{x})\ddot{\hat{\varphi}}(\eta,\xi,\vec{x})\Big].
\end{align}
Recall, that the field operator has the following form:
\begin{align*}
    \hat{\varphi}(\eta,\xi,\vec{x}) =\int_{-\infty}^{+\infty} \frac{d^2k}{2\pi} \int_{0}^{+\infty} \frac{d \omega}{\pi}\sqrt{\sinh\pi \omega}\bigg[e^{-i\omega \eta+i\vec{k}\vec{z}}\hat{b}_{\omega,\vec{k}}^{}  +e^{i\omega \eta- i\vec{k}\vec{z}}\hat{b}_{\omega,\vec{k}}^\dagger\bigg]K_{i\omega}\big(\sqrt{m^2+k^2} e^{\xi}\big).
\end{align*}
Firstly, integration of $\vec{z}$ gives:
\begin{align}
\int_{-\infty}^{+\infty}  d^2\vec{z} e^{i\vec{z}(\vec{k}-\vec{k}')}=(2\pi)^2\delta(\vec{k}-\vec{k}'),
\end{align}
at the same time \cite{PASSIAN2009380}:
\begin{align}
\label{intofK}
    \int_{-\infty}^{+\infty}d\xi K_{i\w}\Big(\sqrt{m^2+k^2}e^\xi\Big)K_{i\w'}\Big(\sqrt{m^2+k^2}e^\xi\Big)=\frac{\pi^2}{2}\frac{\delta(\w-\w')}{\w \sinh \pi \w}.
\end{align}
Thus:
\begin{align}
    \hat{H}=\frac{1}{2}\int_{-\infty}^{+\infty} d^2k \int_0^\infty d\w \w\Big(\hat{b}_{\omega,\vec{k}}^{\dagger} \hat{b}_{\omega,\vec{k}}^{} +\hat{b}_{\omega,\vec{k}}^{} \hat{b}_{\omega,\vec{k}}^{\dagger} \Big).
\end{align}
We use normal ordered Hamiltonian in the density matrix:
\begin{align}
\label{hamrin}
    :\hat{H}:=\int_{-\infty}^{+\infty} d^2k \int_0^\infty d\w \w \ \hat{b}_{\omega,\vec{k}}^{\dagger} \hat{b}_{\omega,\vec{k}}^{}.
\end{align}
Situation in the Schwarzschild space-time is trickier, because we do not know exact expression for the modes. Thus, one can't find formulas similar to \eqref{intofK}. But, in fact, it is enough to know asymptomatic behaviour of the modes. Consider, say, out-going modes:
\begin{align}
\label{intinham}
    I_{\w,\w'}=\int_{-\infty}^{+\infty}dr^*R_{\w,l}(r^*)R^*_{\w',l}(r^*).
\end{align}
These modes solve the following equations (see \eqref{Heq}):
\begin{align}
    \Big[-\partial_{r^*}^2+V_l(r)\Big]R_{\w,l}(r^*)=\w^2 R_{\w,l}(r^*), \qquad \Big[-\partial_{r^*}^2+V_l(r)\Big]R^*_{\w',l}(r^*)=\w'^2 R^*_{\w',l}(r^*).
\end{align}
From these two equations one obtain:
\begin{align}
     R_{\w,l}(r^*)R^*_{\w',l}(r^*)=\frac{1}{\w^2-\w'^2}\partial_{r^*}\Big[R_{\w,l}(r^*)\partial_{r^*}R^*_{\w',l}(r^*)-R^*_{\w',l}(r^*)\partial_{r^*}R_{\w,l}(r^*)\Big].
\end{align}
Thus \eqref{intinham} can be expressed via the behaviour of the modes at the infinities:
\begin{align}
    I_{\w,\w'}=\frac{1}{\w^2-\w'^2}\Big[R_{\w,l}(r^*)\partial_{r^*}R^*_{\w',l}(r^*)-R^*_{\w',l}(r^*)\partial_{r^*}R_{\w,l}(r^*)\Big]\bigg|_{-\infty}^{+\infty}.
\end{align}
Outgoing modes are defined as follows:
\begin{align}
    R_{\w,l}(r^*\to+\infty)\approx T_\w e^{i\sqrt{\w^2-m^2}r^*}, \quad  R_{\w,l}(r^*\to-\infty)\approx e^{i\w r^*}+R_\w e^{-i\w r^*},
\end{align}
where $T_\w $ and $R_\w $ are transmission and reflection coefficients respectively. Normalization coefficient $\frac{1}{\sqrt{2\w}}$ is taken into account in the definition of the field operator \eqref{fieldoperatordesitter}. Using representation of the delta function and Riemann–Lebesgue lemma one can show:
\begin{align}
    \lim_{r^*\to+\infty}\frac{e^{i\alpha r^*}}{\alpha}= \lim_{r^*\to+\infty}\frac{\cos(\alpha r^*)}{\alpha}+i\frac{\sin(\alpha r^*)}{\alpha}=i\pi \delta(\alpha).
\end{align}
Then, straightforward calculation gives:
\begin{align}
     I_{\w,\w'}=\pi \delta(\w-\w')\Big(1+|R_\w|^2+|T_\w|^2\frac{\sqrt{\w^2-m^2}}{\w}\Big).
\end{align}
Transmission and reflection coefficients obey the following condition:
\begin{align}
    \sqrt{\w^2-m^2}|T_\w|^2=\w(1-|R_\w|^2),
\end{align}
finally:
\begin{align}
    I_{\w,\w'}=2\pi \delta(\w-\w').
\end{align}
And the same results may be obtained for the in-going modes. 

To obtain Hamiltonian we use the field operator \eqref{fieldoperatorbh}:
\begin{align}
    \hat{\varphi}(t,r,\phi,\theta)=\sum_{l=0}^{\infty} \sum_{m=-l}^{l}\int_0^\infty d\w\frac{1}{r}\frac{1}{\sqrt{4\pi \w}}  \,\Big[ Y_{l}^m(\phi,\theta)e^{i\w t}\Big(R_{\w,l}\hat{a}^\dagger_{\w,l}+L_{\w,l}\hat{b}^\dagger_{\w,l}\Big)+h.c.\Big].
\end{align}
Here $h.c.$ is a Hermitian conjugated term. Straightforward calculation gives the following Hamiltonian in the Schwarzschild space time:
\begin{align}
\label{hamsch}
    :\hat{H}:=\sum_{l=0}^{\infty} \sum_{m=-l}^{l} \int_0^\infty d\w \w \big( \hat{b}^\dagger_{\w,l}\hat{b}^{}_{\w,l}+\hat{a}^\dagger_{\w,l}\hat{a}^{}_{\w,l}\big).
\end{align}
In the de Sitter space Hamiltonian has the similar form. But, with only one set operators:
\begin{align}
\label{hamds}
    :\hat{H}:=\sum_{l=0}^{\infty} \sum_{m=-l}^{l} \int_0^\infty d\w \w  \hat{b}^\dagger_{\w,l}\hat{b}^{}_{\w,l}.
\end{align}
\section{Estimation of integrals \label{appB}}
In (\ref{I1}) and (\ref{I2}), we have to calculate the integral of the form:

\begin{align*}
   I_n= \int_0^\infty dk k^n K_{i\w}\Big(\sqrt{m^2+k^2}e^\xi\Big)^2, \qquad \text{as} \quad \xi\to-\infty,
\end{align*}
the MacDonald functions have the following integral representation:
\begin{align*}
    K_\nu(z)=\frac{1}{2}\big(\frac{1}{2}z\big)^\nu \int_0^\infty \exp\Big(-t-\frac{z^2}{4 t}\Big)\frac{dt}{t^{\nu+1}}.
\end{align*}
Hence:
\begin{align*}
    I_n=\frac{1}{4}\int_0^\infty dk k^n \int_0^\infty dt_1 dt_2 \frac{1}{t_1^{i\w+1}t_2^{-i\w+1}} e^{-t_1-\frac{(m^2+k^2)e^{2\xi}}{4t_1}}e^{-t_2-\frac{(m^2+k^2)e^{2\xi}}{4t_2}}.
\end{align*}
In the limit $m^2\ll e^{2\xi}$, such an integral simplifies to:
\begin{align*}
    I_n\approx\frac{1}{4}2^n \Gamma\Big(\frac{1+n}{2}\Big) e^{-\xi(n+1)} \int_0^\infty dt_1 dt_2\frac{ (t_1 t_2)^\frac{n-1}{2}}{\big(\frac{t_1}{t_2}\big)^{i\w}(t_1+t_2)^{\frac12(n+1)}}.
\end{align*}
Then we obtain that:
\begin{align*}
    I_1\approx\frac{\pi}{2}\w^2e^{-2\xi}\frac{1}{\w \sinh \pi \w},
\end{align*}
and
\begin{align*}
   I_3\approx \frac{\pi}{3}\Big[\w^2+\w^4\Big]e^{-4\xi}\frac{1}{\w \sinh \pi \w}.
\end{align*}
These results coincide with the expressions found in \eqref{I1} and \eqref{I2} numerically.

\section{Key points of numerical calculations \label{appA}}
\renewcommand\theequation{C.\arabic{equation}}
In this paper, we heavily use the numerical estimates of the sums over $l$ in \eqref{sumsds}, \eqref{sumsbh}, \eqref{sumsbh2}. In this section, we briefly discuss the general aspects of such calculations. Consider, e.g., out-going modes, $R_{\w,l}(r^*)$, in the de Sitter space-time. The equations of motion for these modes are \eqref{Heq}. This is a Schrödinger-like equation for the standard scattering problem. To solve it, we need to fix boundary conditions. For the out-going modes, moving from the horizon to spatial infinity, the proper boundary conditions are of the following form:
\begin{align}
    R^{\text{num}}_{\w,l}(r^*)\propto e^{i \w r^*}, \quad \text{as} \quad r^*\to\infty.
\end{align}
However, we do not know the transmission coefficient, but one can start with, say, the unit coefficient in front of a single wave at plus infinity:
\begin{align}
    R_{\w,l}^0(r^*) = e^{i \w r^*}, \quad \text{as} \quad r^*\to\infty,
\end{align}
then we integrate the equation of motion \eqref{Heq} from some large positive $r_+^*$ to some large negative $r_-^*$. These points have to satisfy the condition $V_l(r_\pm^*)\ll w^2$ because this condition corresponds to the plane-wave-like behavior of the modes. Thus, we obtain two waves at minus infinity, indeed:
\begin{align}
    R_{\w,l}^0(r^*) \approx A e^{i \w r^*} + B e^{-i \w r^*}, \quad \text{as} \quad r^*\to-\infty.
\end{align}
The coefficients $ A, B $ are found by numerical integration. The last step is to set the normalization factor, the amplitude of the wave $e^{i \w r^*}$ at minus infinity should equal to one, then:
\begin{align}
    R^{\text{num}}_{\w,l}(r^*) = \frac{1}{A} R_{\w,l}^0(r^*).
\end{align}
So we obtained the modes $R^{\text{num}}_{\w,l}(r^*)$  numerically. In fact, the sums \eqref{sumsds}, \eqref{sumsbh}, \eqref{sumsbh2} have a natural cut off of $l$ because if at some point $r^*$:
\begin{align}
    V_l(r^*)\gg w^2, \quad\text{then} \quad R^{\text{num}}_{\w,l}(r^*)\ll 1.
\end{align}
\end{appendices}
\bibliographystyle{unsrturl}
\bibliography{bibliography.bib}
\end{document}